\documentstyle[12pt,graphicx]{article}

\begin{document}

\title{Influence of the Nucleon Hard Partons Distribution on $J/\Psi$ Suppression in a GMC Framework\thanks
{Supported by Natural Science Foundation of Hebei
Province(A2008000421)}} \small{\date{}}
\author{Wang Hong-Min$^1$\footnote{E-mail:whmw@sina.com},
   Hou Zhao-Yu$^{2}$, Sun Xian-Jing$^{3}$\\
$^1$\small{Physics Department, Academy of Armored Forces Engineering
of PLA, Beijing 100072, China}\\
$^{2}$\small{Physics Graduate School, Shijiazhuang Railway
Institute, 050043, China}\\
 $^{3}$\small{Institute of High Energy
Physics, Chinese Academy of Sciences, Beijing 100049, China}}
 \maketitle
{\textbf{Abstract:}  In a Glauber Monte Carlo framework, taking
account of the transverse spatial distribution of hard partons in
the nucleon, we analyse the nuclear modification factor $R_{dAu}$
for $J/\psi$ in d+Au collisions with the EPS09 shadowing
parametrization. After the influence of nucleon hard partons
distribution is considered, a clearly upward correction is revealed
for the dependence of $R_{dAu}$ on $N_{coll}$ in peripheral d+Au
collisions, however, an unconspicuous correction is shown for the
results versus $p_{T}$.
The theoretical results are in good agreement with the experimental data from PHENIX.}\\


For $J/\psi$ production is sensitive to both cold nuclear matter
(CNM) and Quark-Gluon Plasma (hot-dense matter)$^{[1]}$, the
\textquotedblleft normal suppression\textquotedblright  in the CNM
should be subtracted in any interpretation of charmonium production
in heavy ion collisions. The Glauber Monte Carlo (GMC)
approach$^{[2-6]}$, which can simulate experimentally observable
quantities and analyse real data, is an ideal tool to study the CNM
effects provided in deuteron-gold (d+Au) collisions$^{[7]}$. In the
framework of the GMC, only the essential $J/\psi$ production
process$^{[8]}$, $g+g\rightarrow c\bar{c}$, is considered. Thus, the
subprocess cross section can be obtained by extracting from the
proton-proton (p+p) experimental data$^{[9]}$ and given as the Monte
Carlo inputs of rapidity $(y)$ and transverse momentum
($p_{T}$)$^{[3,4]}$. The shadowing and \textquotedblleft normal
nuclear absorption\textquotedblright
 effects, which are two dominant CNM effects at Relativistic Heavy Ion Collider (RHIC) and Large Hadron Collider (LHC) energies,
can also be considered properly in the GMC framework. Since
shadowing should depend on the spatial position of the interaction
parton within the nucleus, the inhomogeneous shadowing
effect$^{[10,11]}$ should also be taken into account. In this paper,
the latest shadowing parametrization EPS09 (Eskola, Paukkunen and
Salgado)$^{[12]}$ is used.

At ultra-high energy domain, the mechanism of inelastic hadronic
collisions is dominated by the contribution from small-$x$ gluons
and the influence of the transverse spatial distribution of hard
partons in the nucleon become important$^{[13,14]}$. Unfortunately,
comparing with the longitudinal momentum distribution of partons in
the nucleon, the measurements of the transverse spatial distribution
of partons are rather limited$^{[15]}$. In this paper, two kinds of
hard partons distributions in the nucleon are used. One is assumed
that the hard partons are uniformly distributed in a hard sphere
nucleon, the other is derived from fits to $J/\psi$ photo-production
data at HERA and FNAL$^{[13,15]}$. The corresponding results with
both of them will be given in the following part of the paper.

In the Glauber Monte Carlo framework$^{[2-4]}$, the $J/\psi$
production in nucleus-nucleus (A+B) collisions can be simply written
as
$$\frac{dN^{J/\psi}_{A+B}}{d^{2}\vec{b}dp_{T}}=\frac{1}{N_{r}}\sum_{i=1}^{N_{r}}
\sigma_{nn}^{in}F_{g}^{A}(x_{1i},Q^{2},\vec{b}_{Ai},z_{Ai})F_{g}^{B}(x_{2i},Q^{2},|\vec{b}-\vec{b}_{Ai}|,z_{Bi})$$
\begin{equation}
\times
S_{A}(\vec{b}_{Ai},z_{Ai})S_{B}(|\vec{b}-\vec{b}_{Ai}|,z_{Bi})\frac{dN^{
J/\psi}_{p+p}}{dp_{T}},
\end{equation}
where
$$x_{1}=\frac{m_{T}}{\sqrt{s}}e^{y},~~~~
 x_{2}=\frac{m_{T}}{\sqrt{s}}e^{-y},$$
with the transverse mass of $J/\psi$,
$m_{T}=Q=\sqrt{(2m_{c})^{2}+p_{T}^{2}}$. The charm quark mass
$m_{c}=1.2$GeV and the center of mass energy per nucleon pair
$\sqrt{s}=200$GeV at RHIC energies. $\sigma_{nn}^{in}$ is the
inelastic cross section, $N_{r}$ is the total number of Monte Carlo
random point, and $b$ is the nucleus-nucleus impact
parameter$^{[16]}$. The probability density function for $y$ and
$p_{T}$, which are extracted from p+p data taken from
PHENIX$^{[9]}$, are given by the double Gaussian form and
$C_{1}\times (1+(p_{T}/C_{2})^{2})^{-6}$ form$^{[9,17]}$,
respectively.

 In Eq.(1), the
factor $F_{g}^{A}(x_{1},Q^{2},\vec{b}_{A},z_{A})$ can be given
by$^{[10,11]}$
\begin{equation}
F_{g}^{A}(x_{1},Q^{2},\vec{b}_{A},z_{A})=\rho_{A}(\vec{b}_{A},z_{A})R^{A}_{g}(\vec{b}_{A},x_{1},Q^{2}),
\end{equation}
where $\vec{b}_{A}$ and $z_{A}$ are the transverse and longitudinal
location of the parton in nucleus $A$. If we assume that shadowing
is proportional to the parton path through the nucleus$^{[10,18]}$,
then the inhomogeneous shadowing ratio
\begin{equation}
R^{A}_{g}(\vec{b}_{A},x_{1},Q^{2})=1+N_{\rho}(R_{g}(x_{1},Q^{2})-1){T_{A}(\vec{b}_{A})}/{T_{A}(0)},
\end{equation}
where $T_{A}(\vec{b}_{A})$ is the nuclear thickness function at
$\vec{b}_{A}$, and $R_{g}(x_{1},Q^{2})$ is the homogeneous shadowing
ratio$^{[19]}$ taken from the EPS09 shadowing
parametrization$^{[12]}$. The nucleon density in the nucleus,
$\rho_{A(B)}$, is assumed to be a Woods-Saxon distribution$^{[20]}$
for gold ($Au$) and a H\'{u}lthen form$^{[21,22]}$ for deuteron
($d$). The unitary factor, $N_{\rho}$, is chosen so that $\int
d^{2}\vec{b}_{A}dz_{A}\rho_{A}(\vec{b}_{A},z_{A})
R_{g}^{A}=R_{g}(x_{1},Q^{2})$.

 The survival probability
\begin{equation}
S_{A}(\vec{b}_{A},z_{A})=\textmd{exp}\{-N_{tr}^{J/\psi-N}(\vec{b}_{A},z_{A})\},
\end{equation}
where the number of $c\bar{c}$ colliding with the remaining nucleons
in nucleus $A$$^{[4]}$,
$N_{tr}^{J/\psi-N}=\frac{1}{N_{r}^{\prime}}\sum_{i'}^{N_{r}^{\prime}}\sigma_{abs}$
$\times\rho_{A}(\vec{b}_{A},z^{\prime}_{Aui'}),$ with
$z^{\prime}_{Ai'}>z_{A}$. The absorption cross section,
$\sigma_{abs}$, is taken as $3.1$mb by a global $\chi^{2}$ analysis
with the experimental data from PHENIX$^{[4,7]}$. Both the shadowing
and nuclear absorption effects are ignored for deuteron.

Now let us consider the influence of the transverse spatial
distribution of hard partons in the nucleon. Two kinds of nucleon
hard partons transverse distributions are used in this paper. The
first is assumed that the hard partons are uniformly distributed in
a hard-sphere nucleon, then the distribution function of hard
partons in the transverse plane will be given by
\begin{equation}
F_{1}(b_{n})=3/(2\pi r_{n}^{3})\sqrt{r_{n}^{2}-b_{n}^{2}}
~\theta(r_{n}-b_{n}),
\end{equation}
where the nucleon radius, $r_{n}=\sqrt{\sigma_{nn}^{in}/\pi}/2$, and
the inelastic cross section $\sigma_{nn}^{in}=42$mb (RHIC energies),
$72$mb (LHC energies). The second is derived from the $J/\psi$
photo-production data and described by a dipole form$^{[13]}$
\begin{equation}
F_{2}(b_{n})=m_{g}^{2}/(4\pi)(m_{g}b_{n})K_{1}(m_{g}b_{n}),
\end{equation}
where $K_{1}$ denotes the modified Bessel function$^{[23]}$ and the
mass parameter $m_{g}^{2}\sim 1.1$GeV$^{2}$.

In the GMC framework, the number of binary nucleon-nucleon (n+n)
collisions, $N_{coll}$, is always simply given by$^{[2]}$
\begin{equation}
N_{coll}(b)=\sum_{i\in A,j\in
B}\theta(|\vec{b}_{Ai}-\vec{b}_{Bj}|-d_{max}),
\end{equation}
where $d_{max}=\sqrt{\sigma_{nn}^{in}/\pi}$. If we consider the
nucleon hard partons transverse spatial distribution, the number of
binary n+n collisions will be correspondingly written as$^{[22,24]}$
\begin{equation}
N^{nps}_{coll}(b)=\sum_{i\in A,j\in
B}t_{nps}(\vec{b}_{Ai}-\vec{b}_{Bj})\sigma_{nn}^{in},
\end{equation}
where the normalized overlap function for n+n collision, $t_{nps}$,
can be given by the convolution of collision nucleon hard partons
transverse distribution function. The ratios of $N_{coll}^{nps}$ to
$N_{coll}$ calculated for d+Au collisions are given in Fig.1. In
peripheral collision domain ($b_{dAu}>6$fm), the ratios are less
than 1 and a clearly downward trend can be seen. For the magnitude
of the bias is sensitive to width of the n+n overlap function, the
bias is larger at LHC energies comparing with RHIC energies.

In order to compare with the experimental data from PHENIX, we
introduce the nuclear modification factor:
\begin{equation}
R_{dAu}(b)=\frac{\langle
dN^{J/\psi}_{d+Au}/d^{2}\vec{b}dp_{T}\rangle_{p_{T}}}{N_{coll}(b)\langle
dN_{p+p}^{J/\psi}/dp_{T}\rangle}.
\end{equation}
The $J/\psi$ ratios $R_{dAu}$ versus $N_{coll}$ (left) and $p_{T}$
(right) are shown in Fig.2. From up to down, the results are for
three rapidity regions: backward ($-2.2<y<-1.2$), central
($|y|<0.35$) and forward ($1.2<y<2.2$). The solid curves are the
results without considering the nucleon hard partons distribution.
The dashed and dotted curves are the results with hard-sphere and
dipole nucleon hard partons transverse distribution, respectively.
In the left side of Fig.2, a clearly upward correction is shown at
small $N_{coll}$ for the ratios shown in Fig.1 is less than 1 as
$b_{dAu}>6fm$. It is shown that the theoretical results considered
the nucleon hard partons distribution are in good agreement with the
experimental data from PHENIX$^{[7]}$. In the right side, there
aren't any obvious correction for the ratios versus $p_{T}$. The
reason is that the nuclear modification factor versus $p_{T}$ will
be changed into
\begin{equation}
R_{dAu}(p_{T})=\frac{\langle
dN^{J/\psi}_{d+Au}/d^{2}\vec{b}dp_{T}\rangle_{\vec{b}}}{\langle
N_{coll}(b)\rangle dN_{p+p}^{J/\psi}/dp_{T}},
\end{equation}
and the average number of inelastic n+n collisions at RHIC energies,
$\langle N_{coll}(b)\rangle\sim 1654,$ for all of the methods
mentioned above.

In summary, we have considered the influence of nucleon hard partons
distribution on the nuclear modification factor for $J/\psi$ in d+Au
collisions in the GMC framework. A visible correction can be seen
for the ratios versus $N_{coll}$ and the theoretical results
considered the hard partons distribution are in good agreement with
the experimental data from PHENIX$^{[7]}$. Since the bias shown in
Fig.1 is much larger at LHC energies than RHIC energies with the
same n+n overlap function, the influence of nucleon hard partons
distribution must also be properly considered at LHC energies.

\begin{newpage}

\end{newpage}

\begin{newpage}

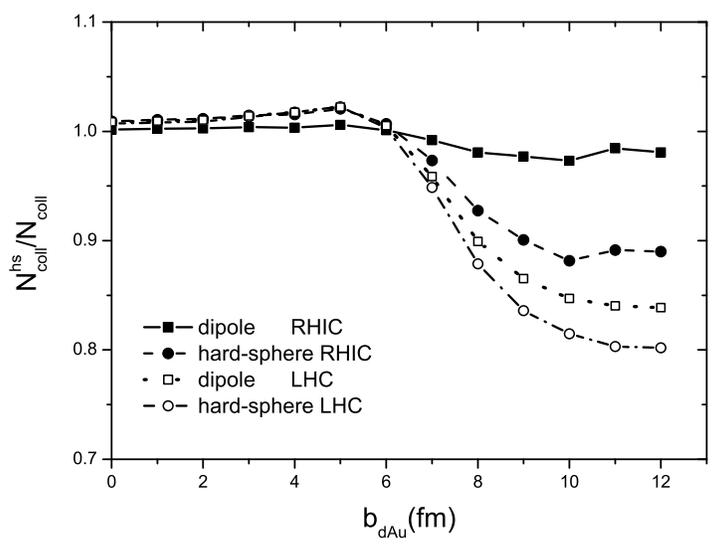
\begin{figure}
\centering  \caption{The ratios of $N_{coll}^{nps}$ to $N_{coll}$
for d+Au collisions at RHIC and LHC energies.}
\end{figure}

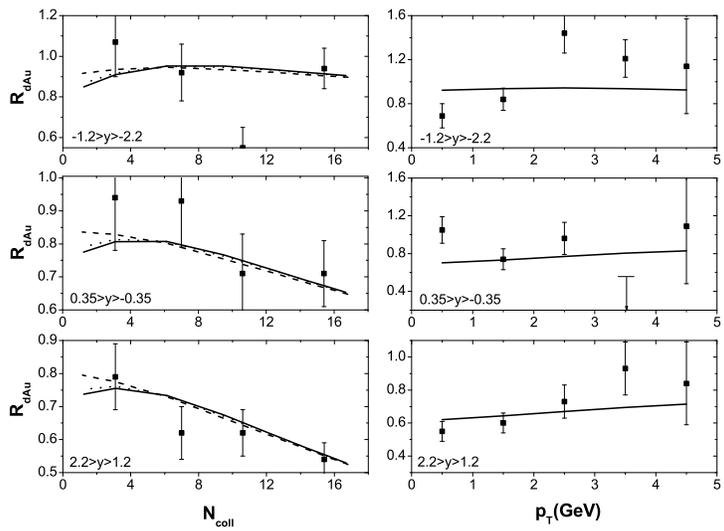
\begin{figure}
\centering  \caption{The $R_{dAu}$ ratios versus $N_{coll}$ (left)
and  $p_{T}$ (right) at RHIC energies for three rapidity ranges:
$-2.2<y<-1.2,~ |y|<0.35,~ 1.2<y<2.2$ (from up to down). The curves
are without considering the nucleon hard partons distribution
(solid), and with hard-sphere (dashed) or dipole (dotted) nucleon
hard partons distribution. The experimental data come from
PHENIX$^{[7]}$.}
\end{figure}

\end{newpage}

\end{document}